\documentclass[fleqn,10pt]{wlscirep}

\usepackage{amssymb} \usepackage{graphicx} \usepackage{amsmath}
\usepackage[T1]{fontenc} \usepackage{pstricks} \usepackage{subfigure}
\usepackage{hyperref}
\usepackage[normalem]{ulem}

\begin{document}

\title{Mechanism for sound dissipation in a two-dimensional degenerate Fermi gas}

\author[1,*]{Krzysztof Gawryluk}
\author[1]{Miros{\l}aw Brewczyk}
\affil[1]{Wydzia{\l} Fizyki, Uniwersytet w Bia{\l}ymstoku, ul. K. Cio{\l}kowskiego 1L, PL-15245 Bia{\l}ystok, Poland}
\affil[*]{k.gawryluk@uwb.edu.pl}

\begin{abstract}
We numerically study the transport properties of a two-dimensional Fermi gas in a weakly and strongly interacting regimes, in the range of temperatures close to the transition to a superfluid phase. For that we excite sound waves in a fermionic mixture by using the phase imprinting technique, follow their evolution, and finally determine both their speed and attenuation. Our formalism, originated from a density-functional theory, incorporates thermal fluctuations via the grand canonical ensemble description and with the help of Metropolis algoritm. From numerical simulations we extract temperature dependence of the sound velocity and diffusivity as well as the dependence on the interaction strength. We emphasize the role of virtual vortex-antivortex pairs creation in the process of sound dissipation.

\end{abstract}
\maketitle

\section{Introduction}

Exciting sound waves within a substance and studying their propagation allows for the exploration of its equilibrium and dynamical properties. The characteristics like the compressibility and viscosity become available via measuring the speed of sound waves and their attenuation.

The propagation of sound in harmonically trapped and homogeneous systems of ultracold bosonic and fermionic atomic gases has been studied experimentally by many groups \cite{Ketterle1997,Hoefer06,Joseph07,Chang08,Meppelink09,Sidorenkov13,Ville18,Patel20,Bohlen20,Hadzibabic21,Patel23}. In a recent work on a uniform two-dimensional weakly interacting Bose gas a sound wave, moving with velocity close to the Bogoliubov sound speed, was observed below and above the critical temperature for the Berezinskii-Kosterlitz-Thouless (BKT) transition \cite{Ville18}. This phenomenon has been investigated and verified by various theoretical papers \cite{Ota18,Cappellaro18,Singh20,Wu20,Gawryluk21}. In Ref. \cite{Hadzibabic21}, the first and the second sound modes were observed for the first time in 2D Bose gas. Based on temperature dependence of sound speeds the superfluid density of bosonic gas was determined. The superfluid density revealed a universal jump at the critical temperature, in accordance with the BKT theory and as already proved numerically for a uniform two-dimensional atomic Bose gas in Ref. \cite{Gawryluk19}. A single density wave was observed also in interacting atomic Fermi gas at temperatures below the superfluid transition temperature \cite{Bohlen20} (for a theoretical description see Ref. \cite{Tononi21}). The damping of the sound mode was measured and it was found that the damping is minimal in the strongly interacting regime. For the strongly interacting regime the diffusivity approaches the universal quantum limit which is $\hbar/m$, where $m$ is the mass of an atom. The results of Ref. \cite{Bohlen20} demonstrate that the  strongly interacting regime distinguishes from the BCS/BEC sides of the BEC-BCS crossover with respect to the dissipation properties. The experimental data presented in \cite{Patel20} show that the diffusivity achieves a universal quantum limit also in a three-dimensional homogeneous atomic Fermi gas at the unitarity and below the superfluid transition temperature. When the temperature gets higher the diffusivity increases monotonically.

In this paper we focus on the sound propagation in two-dimensional Fermi-Fermi mixtures at low temperatures (but still above the superfluid transition temperature) in a weakly and strongly interacting regimes. Several approaches to study out-of-equilibrium bosonic and fermionic quantum many-body dynamics has been already developed, including those based on the classical field \cite{Brewczyk07,Blakie08,Proukakis08} and the two-particle irreducible effective action, the renormalisation-group theory, or the Hartree-Fock-Bogoliubov \cite{Gasenzer09,Gasenzer11,Pawlowski21} approximation. Here, we demonstrate yet another attempt to fermionic systems based on density-functional theory in its time-dependent version, which has regard to the thermal fluctuations (for a density-functional theory version valid in a fermionic superfluid phase see Ref. \cite{Barresi23}). It can be viewed as a continuation of our previous studies of thermal fermionic systems \cite{Grochowski20,Ryszkiewicz22}. In our approach, after excitation, we can follow sound waves propagation and have an access both to their speed and the damping related properties. The damping turns out to be minimal for the values of the interaction parameter close to zero and the diffusivity gets close to the universal quantum limit. Opposite is true for positive and negative values of the interaction parameter, where the diffusivity becomes very large.

The paper is organized as follows. In Sec. \ref{Method} we introduce the model of a two-component two-dimensional degenerate Fermi gas at temperatures above the transition to a superfluid phase. Then (Sec. \ref{soundwaves}) we describe the numerical experiment in which we excite the sound waves in the system, allow them to propagate, and at the end determine their speed and attenuation. In Sec. \ref{Mechanism} we reveal the mechanism of sound dissipation which turns out to be related to creation of virtual vortex-antivortex pairs in the gas. Finally, we conclude in Sec. \ref{conclusion}.

\section{Method}
\label{Method}
We start with a simple description of a two-component fermionic mixture (a mixture of atoms being in two internal degrees of freedom as in experiment of Ref. \cite{Bohlen20}) in terms of semiclassical distribution functions, $f_{\bf{p}}^{\pm}({\bf{r}})$, already detailed in Ref. \cite{Ryszkiewicz22}. Here, indices $"+"$ and $"-"$ distinguish components and the equilibrium, at a given temperature $T$ and a chemical potential $\mu$, semiclassical distributions are $f_{\bf{p}}^{\pm}({\bf{r}})=(\exp{[\varepsilon_{\bf{p}}^{\pm}({\bf{r}})-\mu_{\pm}]/k_B T}+1)^{-1}$. The particle energy $\varepsilon_{\bf{p}}^{\pm}({\bf{r}})$ at position ${\bf{r}}$ includes the kinetic energy as well as the potential energy related to external trapping and interactions. According to the meaning of semiclassical distribution functions the integrals $\int f_{\bf{p}}^{\pm}({\bf{r}}) d{\bf{p}}$ and $\int ({\bf{p}}^2/2m) f_{\bf{p}}^{\pm}({\bf{r}}) d{\bf{p}}$ represent the atomic densities and atomic local motion (intrinsic) energies in both components. We further assume that we consider the case of a tightly-confined two-dimensional gas in a box-shaped trap. The atomic and the local motion energy densities then are $n_{\pm}({\bf{r}})=(1/\lambda^2) \ln{(1+z_{\pm}({\bf{r}}))}$ and $\varepsilon_{\pm}({\bf{r}})=(k_B T/\lambda^2)\, f_2\,(z_{\pm}({\bf{r}}))$, where $\lambda=\sqrt{2\pi\hbar^2/m k_B T}$ is the thermal wavelength, $z_{\pm}({\bf{r}})=\exp{((\mu_{\pm} - \delta V_{int}/\delta n_{\pm} )/ k_B T)}$ are the extended fugacities, the functional $V_{int}(n_+,n_-)$ describes the interaction between components, and $f_{2}(z)=2 \int_0^{\infty} x^3/(z^{-1} e^{x^2}+1)\, dx$ is one of the standard functions for fermions \cite{Huang}. The free energy functional (which, according to Ref. \cite{Kohn65}, substitutes the energy functional in the case of nonzero temperatures) of a whole system can be written as
\begin{equation}
F_{tot}(n_+,n_-,{\bf{v}}_{+},{\bf{v}}_{-}) = \int \left[ f_{+}({\bf{r}}) + f_{-}({\bf{r}}) \right]\, d{\bf{r}}  
+ \int \left( n_+ \frac{1}{2} m {\bf{v}}_{+}^2 + n_- \frac{1}{2} m {\bf{v}}_{-}^2 \right) d{\bf{r}}
+ V_{int}(n_+,n_-)  \,. 
\label{Ftotal}
\end{equation}
The energy depends on atomic densities, $n_{\pm}({\bf{r}})$, and macroscopic velocity fields, ${\bf{v}}_{\pm}({\bf{r}})$, of both fermionic gases. The density of a local free energy of a two-dimensional gas is
\begin{eqnarray}
f_{\pm}({\bf{r}}) = \frac{k_B T}{\lambda^2} \left[ (\ln{z_{\pm}})\, \ln{(1+z_{\pm})} - f_{2}(z_{\pm}) \right] 
\label{freeenergy2D}
\end{eqnarray}
and the second integral in Eq. (\ref{Ftotal}) represents the energy of a macroscopic flow.

It is convenient to represent the density and velocity fields by a single complex field $\psi_{\pm}({\bf{r}})$ introduced via inverse Madelung transformation \cite{Deb98,Domps98,Grochowski17}. The density and velocity fields are related to the pseudo-wave functions $\psi_{\pm}({\bf{r}})$ as $n_{\pm}=|\psi_{\pm}|^2$ and ${\bf v}_{\pm}=(\hbar/m)\, \nabla \phi_{\pm}$, where $\phi_{\pm}({\bf{r}})$ are the phases of complex functions $\psi_{\pm}({\bf{r}})$. Then the density of a macroscopic flow energy can be expressed in the following way
\begin{eqnarray}
-\frac{\hbar^2}{2m} \psi_{\pm}^*\,  \nabla^2 \psi_{\pm} - \frac{\hbar^2}{2m}  
(\nabla |\psi_{\pm}|)^2  =  n_{\pm} \frac{1}{2}\, m\, {\bf v}_{\pm}^2  
\label{splitting}
\end{eqnarray}
and the functional Eq. (\ref{Ftotal}) is transformed to
\begin{eqnarray}
F_{tot}(\psi_{\pm},\nabla \psi_{\pm}) = \int \left[ f_{+}({\bf{r}}) + f_{-}({\bf{r}}) \right]\, d{\bf{r}}  
+ \int \Big[-\frac{\hbar^2}{2m} \psi_+^* \nabla^2 \psi_+ - \frac{\hbar^2}{2m} (\nabla |\psi_+|)^2 \Big] d{\bf{r}}
\nonumber  \\
+ \int \Big[-\frac{\hbar^2}{2m} \psi_-^* \nabla^2 \psi_- - \frac{\hbar^2}{2m} (\nabla |\psi_-|)^2 \Big] d{\bf{r}}
 +\, V_{int}(n_+,n_-)  \,.
\label{Ftotalnew}
\end{eqnarray}
The equations of motion corresponding to the functional Eq. (\ref{Ftotalnew}) are found in a usual way as 
$i \hbar\, (\partial/\partial t) \psi_{\pm}({\bf r},t) = (\delta/\delta \psi_{\pm}^*) 
F_{tot}(\psi_{\pm}, \nabla \psi_{\pm})$ and are given by
\begin{equation}
i \hbar\, \frac{\partial \psi_{\pm}}{\partial t} =
\left[ -\frac{\hbar^2}{2m} \nabla^2 + \frac{\hbar^2}{2m} \frac{\nabla^2 |\psi_{\pm}|}{|\psi_{\pm}|} +
k_B T\, \ln{z_\pm} 
+ \frac{\delta V_{int}}{\delta n_\pm}  \right] \psi_{\pm} \,.
\label{equmotion2D}
\end{equation}
While evolving pseudo-wave functions according to Eqs. (\ref{equmotion2D}), the extended fugacities $z_\pm({\bf{r}})$ are found from the self-consistency condition $n_{\pm}=(1/\lambda^2) \ln{(1+z_{\pm})}$ with $n_\pm=|\psi_\pm|^2$. 

The interaction between components is treated within the beyond mean-field approach, the lowest-order constrained variational (LOCV) method \cite{Pandharipande73,Pandharipande77,Cowell02,Taylor11,Yu11,Grochowski20}. The interaction energy of a uniform balanced system of density $n$ (per component) is calculated as $E_{int}/V = (\hbar^2/m)\, n^2 A(\eta)$, where $A(\eta)$ is a function of dimensionless parameter $\eta = \ln{(k_F a_{2D})}$ with $k_F=\sqrt{4\pi n}$ being the Fermi momentum -- see Supplementary materials for a derivation based on LOCV approximation and a pseudopotential technique for the two-dimensional contact interactions \cite{Whitehead16}. The numerically determined function $A(\eta)$ for two lowest energy branches (attractive and repulsive ones) is plotted in Fig. \ref{Aeta} (left frame) for the parameter $\eta$ scanning the BCS to BEC crossover. Since the relative population of two lowest branches is $\exp{\left[-(A_2(\eta)-A_1(\eta))\, T_F/(2\pi T)\right]}$, where $T_F$ is the Fermi temperature and $A_1(\eta)$ ($A_2(\eta)$) determines the pair energy in the lower (higher) branch, it is clear that mostly lower branch is occupied for the low temperatures we consider (i.e., $T/T_F<0.2$). 

\begin{figure}[thb] 
\center
\includegraphics[width=7.0cm]{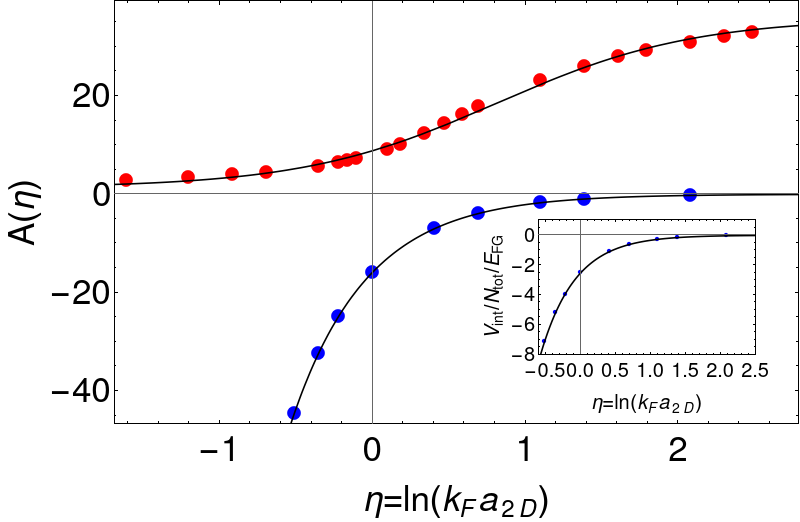}
\includegraphics[width=7.0cm]{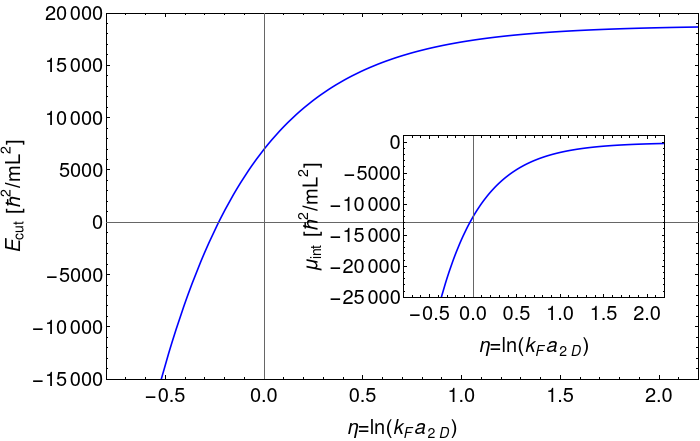}
\caption{Left frame: $A(\eta)$, where $\eta = \ln{(k_F\, a_{2D})}$, calculated within 2D LOCV approximation for two lowest energy branches. The solid black lines are just a guide to the eye. Inset shows the interaction energy per particle in units of the energy per particle of the noninteracting gas, $V_{int}/N_{tot}/E_{FG}$, where $E_{FG}=\pi (\hbar^2/m) n $. The data compare well with the results of the fixed-node diffusion Monte Carlo calculations of Ref. \cite{Bertaina11} (see Fig. 1).  
Right frame: Cutoff energy as a function of $\eta = \ln{(k_F\, a_{2D})}$ for the mixture consisted of $\langle N_{\pm} \rangle =1500$ atoms at the temperature $T/T_F=0.2$. Inset shows the chemical potential $\mu_{int}$. }
\label{Aeta}
\label{cutoffenergy}
\end{figure}

Now we introduce thermal fluctuations into our description. We propose here a different way than the one used in Ref. \cite{Grochowski20}. The grand canonical ensemble of pseudo-fields $\psi_\pm$ fulfilling Eqs. (\ref{equmotion2D}) can be obtained from the free energy functional Eq. (\ref{Ftotalnew}) by using the Metropolis algorithm \cite{Witkowska10,Gawryluk17,Pietraszewicz17}. At a given temperature $T$ and a chemical potential $\mu$, the probability of having a system with $N$ particles and the free energy equal to $F(N)$, calculated within the grand canonical ensemble, is $e^{\mu N /k_B T} e^{-F(N)/k_B T}$. For our system the free energy $F(N)$ is given by Eq. (\ref{Ftotalnew}), where $N$ is the total number of atoms. The fields $\psi_\pm$ are determined by expanding them in a particular set of basis functions. In our case a uniform spatial grid with a given step $\Delta$ is used (i.e. both fields are expanded in a set of Dirac delta functions). The maximal energy available on a grid with spatial step $\Delta$ equals $\hbar^2 (\pi/\Delta)^2/2m$. This, introduced by discretization, cutoff energy should allow to include all energy-relevant many-body states while generating the grand canonical ensemble. For noninteracting mixture of Fermi gases it is meaningful to take the cutoff energy, due to thermal broadening of the Fermi-Dirac distribution, as $E_{cut}=E_F + \alpha\, k_B T$, where $E_F=2\pi (\hbar^2/m)\, n$ is the Fermi energy and $\alpha \gtrsim 1$ (we take $\alpha=5$). For interacting gases we include the interaction-related chemical potential and modify the cutoff as $E_{cut} \rightarrow E_{cut} + \delta V_{int}/\delta n$. Fig. \ref{cutoffenergy}, right frame (details in Suplementary materials) shows the cutoff energy as a function of $\eta = \ln{(k_F\, a_{2D})}$ for the mixture consisted of $\langle N_{\pm} \rangle =1500$ atoms at the temperature $T/T_F=0.2$. 

Fig. \ref{cutoffenergy} (right frame) suggests that our description of fermionic mixture breaks down for negative values of interaction parameter $\eta$ when the system enters the BEC phase of the crossover and the mixture should be described rather in terms of composite bosons (then all treatments of bosonic particles at nonzero temperatures are at hand \cite{Brewczyk07,Blakie08,Proukakis08}). However, an alternative way to achieve the proper cutoff energy exists. Since the total chemical potential is $\mu_{tot}(\eta) \approx E_F + \mu_{int}(\eta)$ (and $\mu_{int}=\delta V_{int}/\delta n$, shown in the inset of Fig. \ref{cutoffenergy}, right frame), we search for the appropriate grid via generating the grand canonical ensemble of pseudo-fields $\psi_\pm$ at the condition $\langle N_{\pm} \rangle =1500$ for each interaction strength. The cutoff energy found in this way remains close to that obtained as described earlier for any interaction strength provided the absolute value $|E_F + \alpha\, k_B T + \delta V_{int}/\delta n|$ is used as the definition for a cutoff energy.

\section{Sound waves propagation}
\label{soundwaves}
\begin{figure}[thb]
\center
\includegraphics[width=7.0cm]{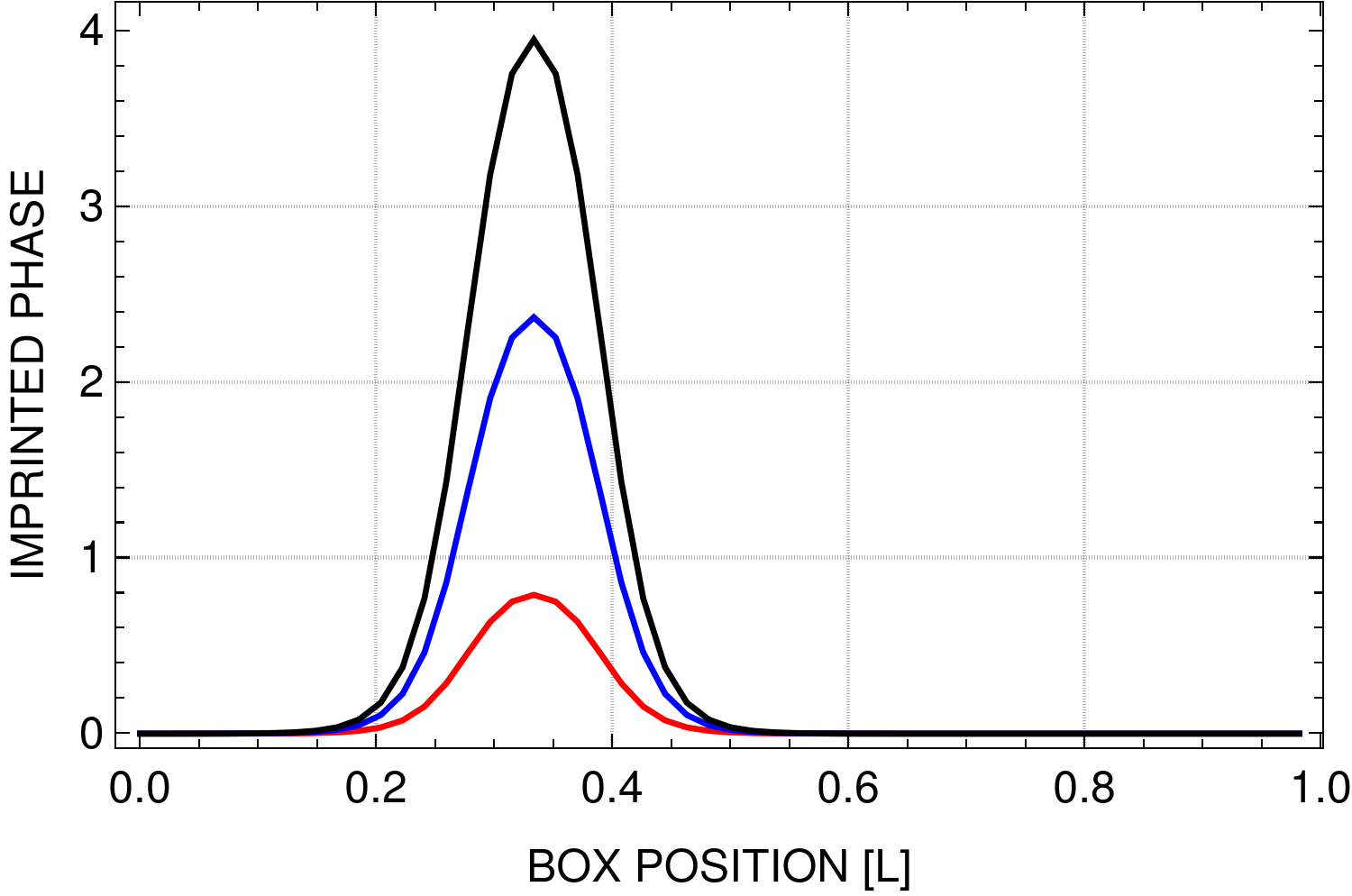}
\caption{Phase profiles used for imprinting procedure applied on pseudo-fields $\psi_{\pm}$. Three curves correspond to different strengths of disturbances: $\phi_0=1.3\, \pi$, $\phi_0=0.8\, \pi$, and $\phi_0=0.27\, \pi$ (from top to bottom).}
\label{phase_imp}
\end{figure}

\begin{figure}[thb]
\center
\includegraphics[width=8.5cm]{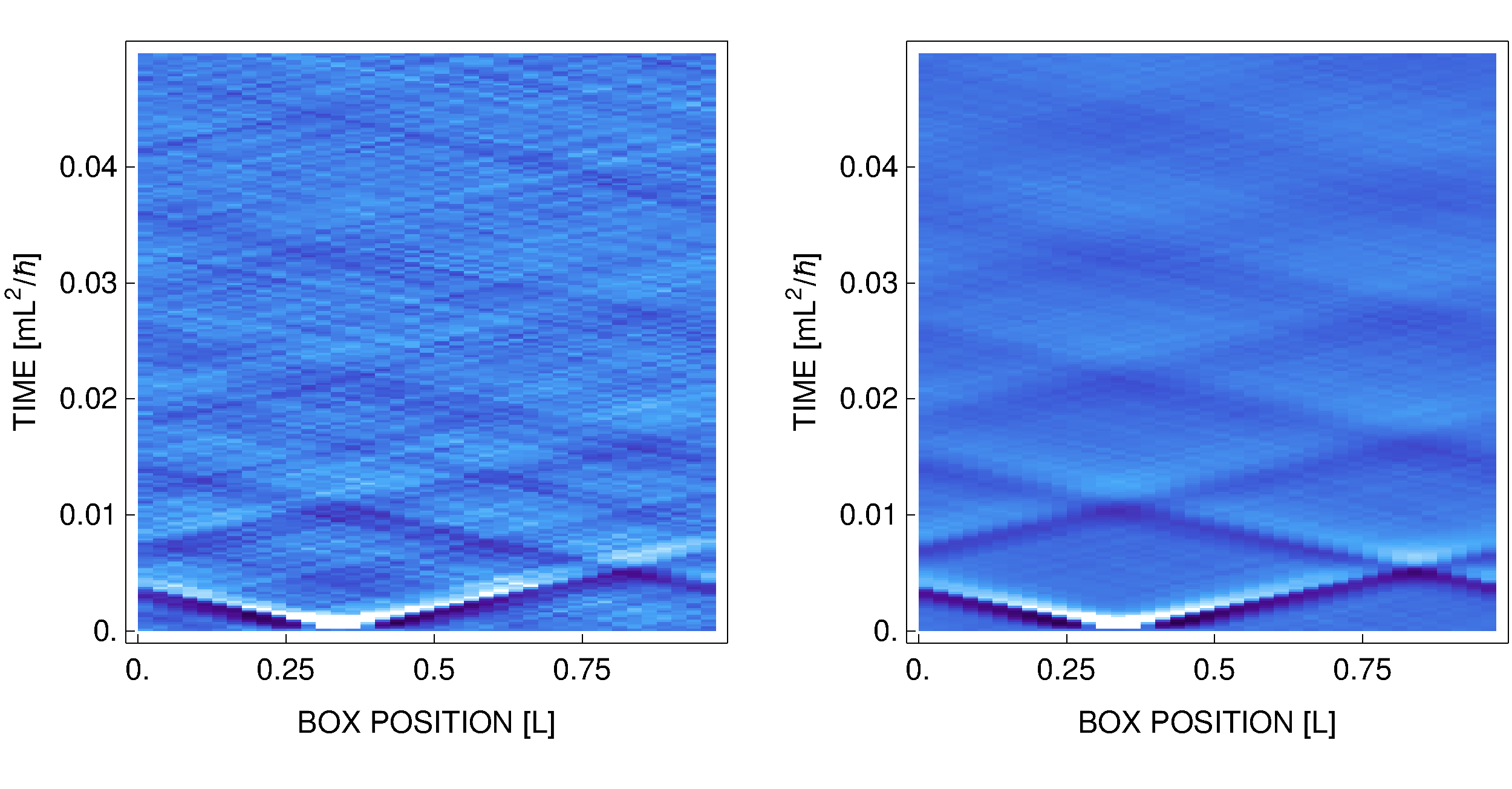}
\caption{Evolution of a density (integrated along $y$ direction) following the phase imprinting. Left (right) picture is an average over $10$ ($200$) realizations (see text). Both frames correspond to the strength of the imprinted phase $\phi_0=0.8\, \pi$, the temperature $T=0.1\,T_F$, and $\eta=0.5$. }
\label{evoDen2D}
\end{figure}

To excite sound waves we disturb the mixture of fermionic atoms with the protocol similar to that applied in experimental work of Ref. \cite{Bohlen20}. Namely, we imprint a phase profile on the pseudo-wave functions $\psi_{\pm}(x,y) \to \psi_{\pm}(x,y) \exp^{i \phi(x)}$ along $x$ direction, where $\phi(x)=\phi_0 \exp{[-(x-0.3\, L)^2/\sigma^2]}$ (see Fig. \ref{phase_imp}). Here, $\phi_0$ determines the strength of the phase disturbance, $L$ is the size of a two-dimensional atomic system, and $\sigma=L/13$ gives the width of the region with imprinted phase. This disturbance results in increase in the total energy of the system $\Delta E=E_{imp}-E_{ini}$, which depends on the particular value of $\phi_0$ and on the temperature $T$. For example, $\Delta E=16\, \hbar^2/m L^2$ for the smallest perturbation $\phi_0=0.27\, \pi$, $\Delta E=28\, \hbar^2/m L^2$ for $\phi_0=0.8\, \pi$, and $\Delta E=50\, \hbar^2/m L^2$ for the strongest perturbation considered $\phi_0=1.3\, \pi$, for $T=0.15\,T_F$ and $\eta=1$ (here, $E_{ini}=2151\, \hbar^2/m L^2$). Note that we use a profile as in Fig. \ref{phase_imp} with a symmetrical, double phase jumps (we work with periodic boundary conditions and the care about phase continuity should be taken). Both phase jumps convert to the density perturbations and result in an appearance of dark and white soliton-like structures traveling in opposite directions, see Refs. \cite{Karpiuk02a,Karpiuk02b}. In our case the density structures are hidden inside a thermal noise, so they are almost not detectable from density profiles. To overcome this problem we repeat the phase imprinting procedure on many grand canonical microstates (typically $200$) and average the density over all realizations. As a result we obtain clear picture of a density response of the system to the initial perturbation. To see density time evolution on a single plot we also integrate two-dimensional density along one spatial coordinate (perpendicular to the direction of sound wave motion) obtaining time-dependent density pictures like in Fig. \ref{evoDen2D}.

Typical density evolution after phase imprinting is shown in Fig. \ref{evoDen2D}. One can see two wave forms propagating in the opposite directions with respect to the region where the phase profile has the maximum. In fact, each of this structures is made of a dark and a white quasisoliton formed from opposite slopes of the phase profile (they are close to each other and almost form a single entity). We could change this behavior by making the width of a phase profile wider, but then we would see in fact four distinguishable structures propagating in the system, which differs from what is seen in the experiment of Ref. \cite{Bohlen20}. Also note that the propagating sound mode does not bounce from the walls since we use periodic boundary condition -- instead the signal is propagating from the other side after reaching the "edge". As in the experiment \cite{Bohlen20}, the initial signal is traveling with constant velocity (we work in a weak disturbance regime) and is damped while propagating, and finally disappears.

\begin{figure}[thb]
\center
\includegraphics[width=16.cm]{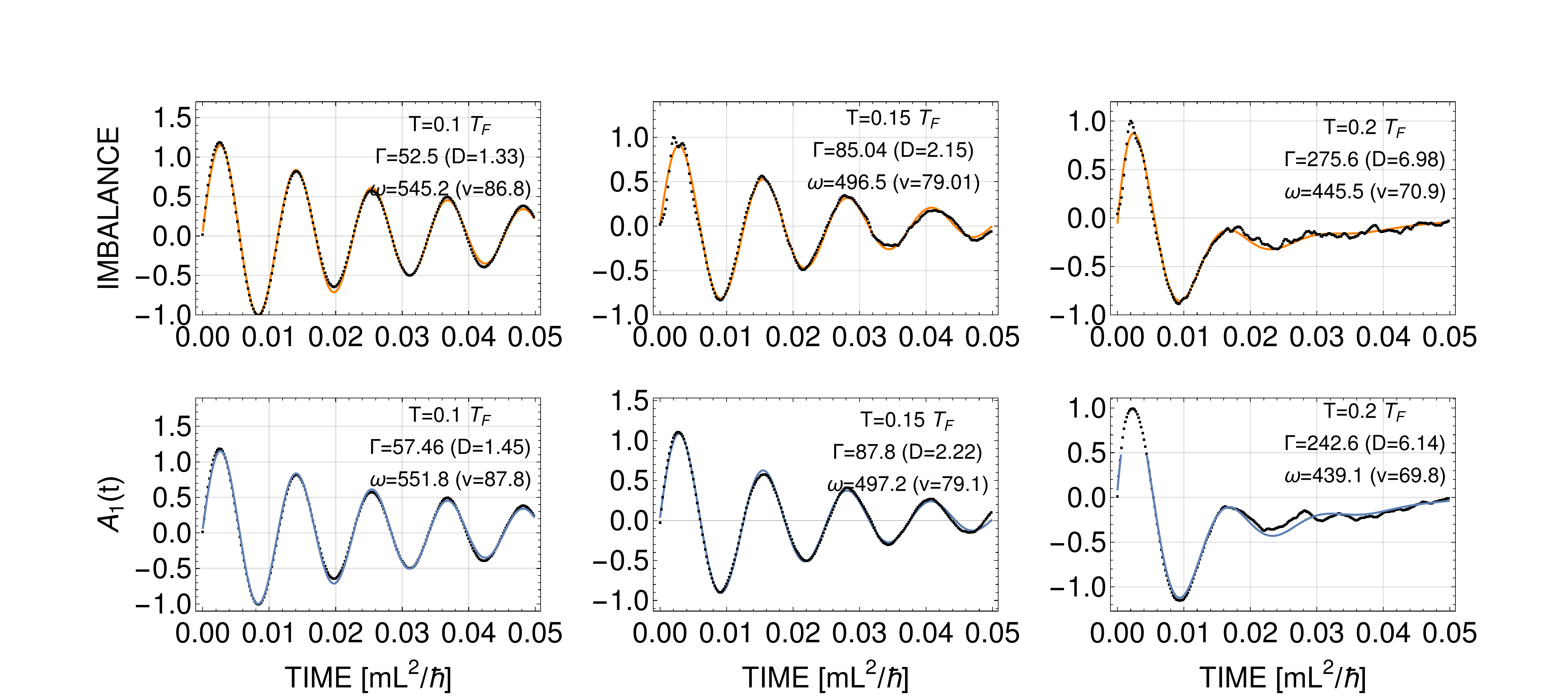}
\caption{Evolution of the density imbalance (top) and the lowest Fourier mode (bottom) after phase imprintings for $\eta=0.5$ and different temperatures: $T/T_F=0.1$ and $\phi_0/\pi=0.27$, $T/T_F=0.15$ and $\phi_0/\pi=0.8$, and $T/T_F=0.2$ and $\phi_0/\pi=1.3$, from left to right. Each picture is an average over $200$ realizations. In each frame the extracted values of the velocity field $v$ (in units of $\hbar/(m L)$) and the diffusivity $D$ (in units of $\hbar/m$) are given. 
}
\label{fitting1}
\end{figure}

To analyze a sound signal we determine its velocity ($v$) and the diffusion coefficient ($D$). For that we calculate the density imbalance defined as $\Delta n(t) = (n_l - n_r)/(n_l + n_r)$, where $n_l$ ($n_r$) is density in the left (right) halves of the potential box, as a function of time, just like in Ref. \cite{Bohlen20}. In the other way we integrate the density along the transverse direction to the propagation and Fourier decompose as $n(x,t)=\langle n \rangle + \sum_{j=\pm 1,\pm 2,...} A_j(t) \exp(j\,2\pi x/L)$, and look at the time-dependence of the Fourier coefficient $A_1(t)$ (in fact, $A_1(t)/{\rm max}[A_1(t)]$) as in Ref. \cite{Ville18}. Then we fit both the density imbalance and the $A_1(t)$ coefficient to an exponentially damped periodic function of the form: $a \exp(-\Gamma t/2)\sin(\omega t + \varphi) + b t+c$ (linear part of the fit is used to neglect influence of constant drift of a signal), see Fig. \ref{fitting1}. We calculate the velocity as $v=L\, \omega/(2\pi)$ and the diffusion coefficient as $D=L^2\, \Gamma/(2\pi)^2$ \cite{Bohlen20}. Generally, both methods give close results, and only in few cases they differ or fail to fit the data (only matching results are accepted).

\begin{figure}[thb]
\center
\includegraphics[width=7.0cm]{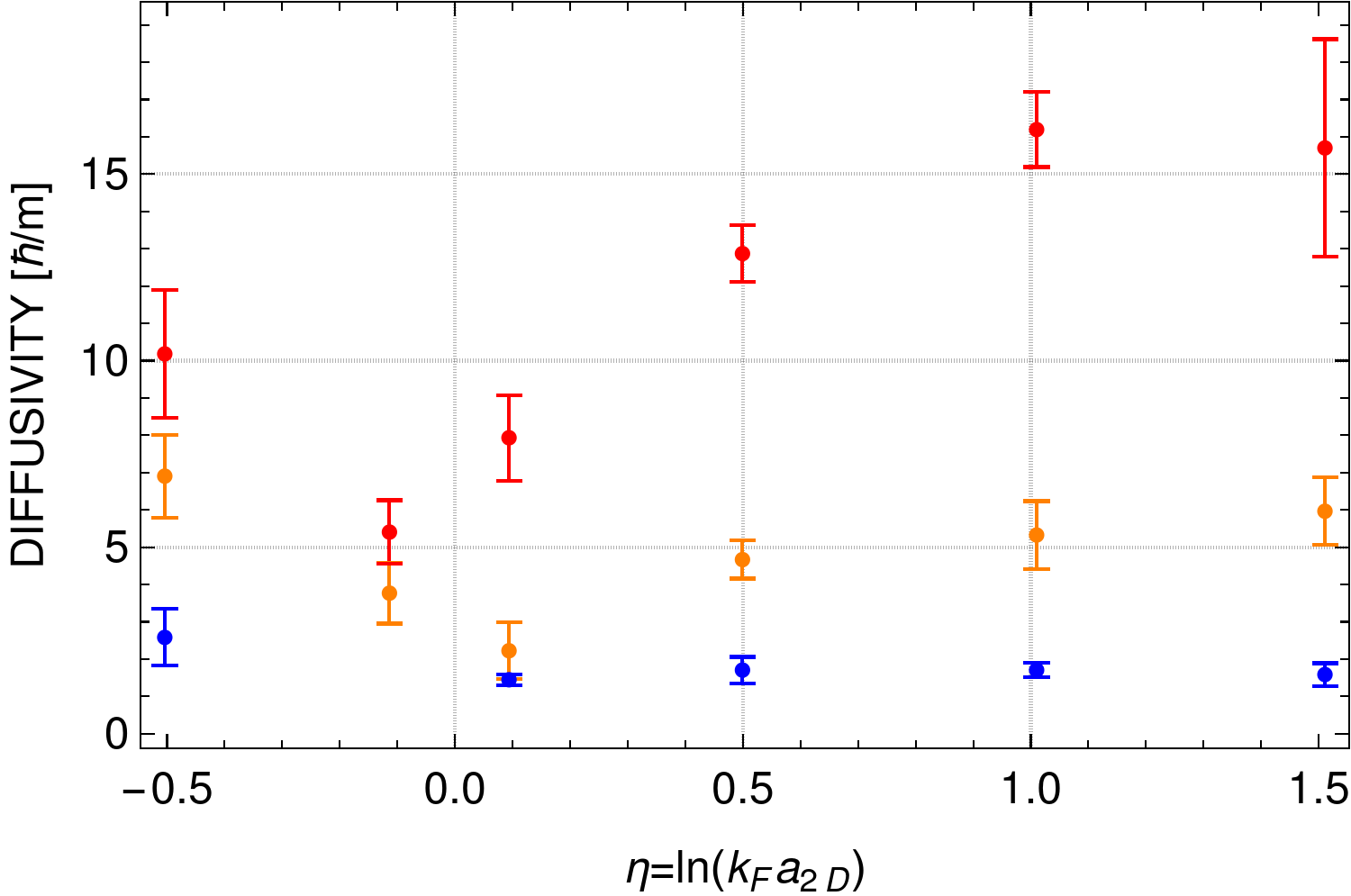}
\caption{Diffusion coefficient as a function of interaction parameter for different temperatures: $T/T_F=0.1$ (blue color), $T/T_F=0.15$ (orange color), and  $T/T_F=0.2$ (red color). Error bars represent the variance calculated based on three different sets of time evolution after imprinting the phase disturbed with $\phi_0/\pi=0.27,0.8,1.3$ for $T/T_F=0.1$, $\phi_0/\pi=0.8,1.3$ for $T/T_F=0.15$, and $\phi_0/\pi=1.3$ for $T/T_F=0.2$.}
\label{figD}
\end{figure}

\begin{figure}[thb]
\center
\includegraphics[width=7.0cm]{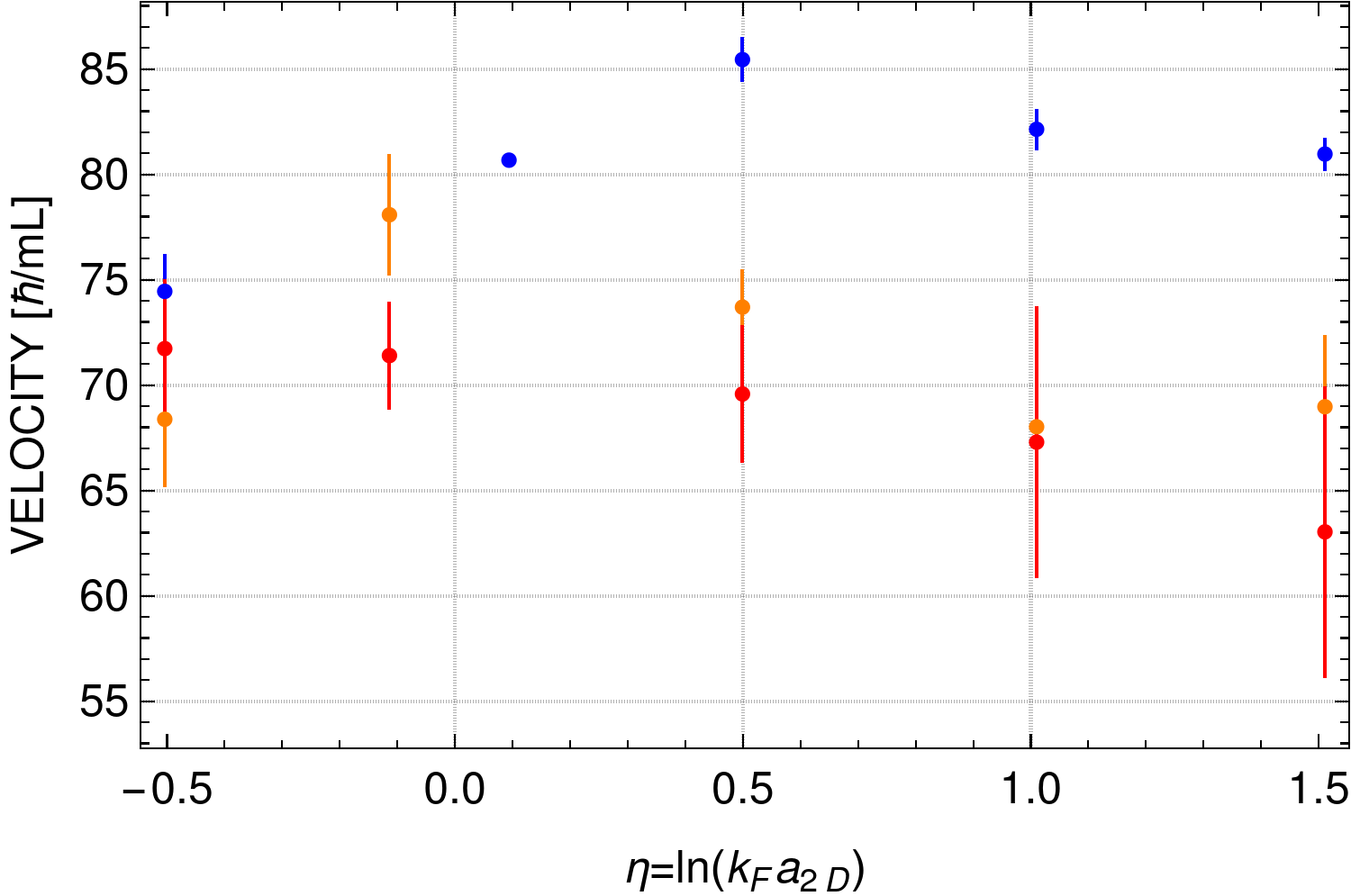}
\caption{Sound velocity as a function of interaction parameter for different temperatures: $T/T_F=0.1$ (blue color), $T/T_F=0.15$ (orange color), and  $T/T_F=0.2$ (red color). Error bars as in Fig. \ref{figD}.}
\label{figV}
\end{figure}

Densities presented in Fig. \ref{evoDen2D} are smooth enough to obtain values of $\Gamma$ and $\omega$, but still these values change depending on the realization, especially for the highest temperature analyzed. This is why we analyzed three sets of real time evolution of the mixture after the phase imprinting, each including $200$ realizations (microstates) from the grand canonical ensemble. This allow us to have more data to present, and add the variance as additional characteristics. Results are summarized in Figs. \ref{figD} and \ref{figV}. Two features of dissipation mechanism are clearly visible. The dissipation increases with temperature. It is true for all considered values of the interaction parameter. Similar observation is reported in \cite{Patel20}, where a three-dimensional Fermi gas is studied at unitarity. In two-dimensional system, however, the diffusivity grows with temperature much faster -- at $T/T_F=0.2$ the diffusivity is already as large as in three-dimensional case for $T/T_F=1$ (see Fig. $4$ in \cite{Patel20}). The reason for that is explained in the next section. The sound dissipation also depends on the interaction strength. It is the weakest in the region of $\eta$ close to zero, when the diffusivity approaches the quantum limit of $\hbar/m$ at the lowest temperature. Similar behavior of diffusivity coefficient was observed in experiment \cite{Bohlen20} performed at temperatures below the superfluid transition temperature. As shown in the next section, the results presented in Figs. \ref{figD} and \ref{figV} can be understood via the dissipation mechanism based on creation of virtual vortex-antivortex pairs.

An estimation of a sound velocity at low temperatures can be done based on Eqs. (\ref{equmotion2D}). The Madelung-representation \cite{Madelung27} equations corresponding to Eqs. (\ref{equmotion2D}), that read (just for one component)
\begin{eqnarray}
&&\frac{\partial n_{+}}{\partial t} + \nabla \cdot (n_+ {\bf v}_+) = 0 \nonumber  \\
&&m \frac{\partial {\bf v}_+}{\partial t} + \nabla \Big( k_B T\, \ln{z_+} + \frac{\delta V_{int}}{\delta n_+} + \frac{1}{2} m {\bf v}_{+}^2 \Big) = 0  \,,  
\label{Madelung}
\end{eqnarray}
can be solved by assuming small deviation of a density from the equilibrium one (which is constant): $n_+=n_{eq}^+ + \delta n_+$. At the low-temperature limit one has $n_+ \lambda^2=\ln{(1+z_+)} \approx \ln{z_+}$, since $z_+ \gg 1$. Then $k_B T\, \ln{z_+} \approx k_B T\, \lambda^2\, n_+ = (2\pi \hbar^2/m)\, n_+$. Leaving only small quantities of the first order, the continuity equation becomes $\partial/\partial t\, \delta n_+ = -n_{eq}^+\, \nabla \cdot{\bf v}_+$. In the weak interaction case this equation can be combined with the equation of motion leading to the following wave equation
\begin{eqnarray}
\frac{\partial^2}{\partial t^2} \delta n_+ - \Big( \frac{2\pi \hbar^2}{m^2}\, n_{eq}^+ \Big)\,  \nabla^2 \delta n_+  = 0 \,.  
\label{Madelung1}
\end{eqnarray}
The expression in the bracket gives the square of the sound velocity. Hence, for the parameters as used for Fig. \ref{figV} (here, $\langle N_+\rangle=1500$) the estimation of the sound velocity is $\sqrt{(2\pi \hbar^2/m^2)\, n_{eq}^+}\, \approx\, 97\, \hbar/(m L)$.

\section{Mechanism for sound dissipation}
\label{Mechanism}

\begin{figure}[thb]
\center
\includegraphics[width=10.0cm]{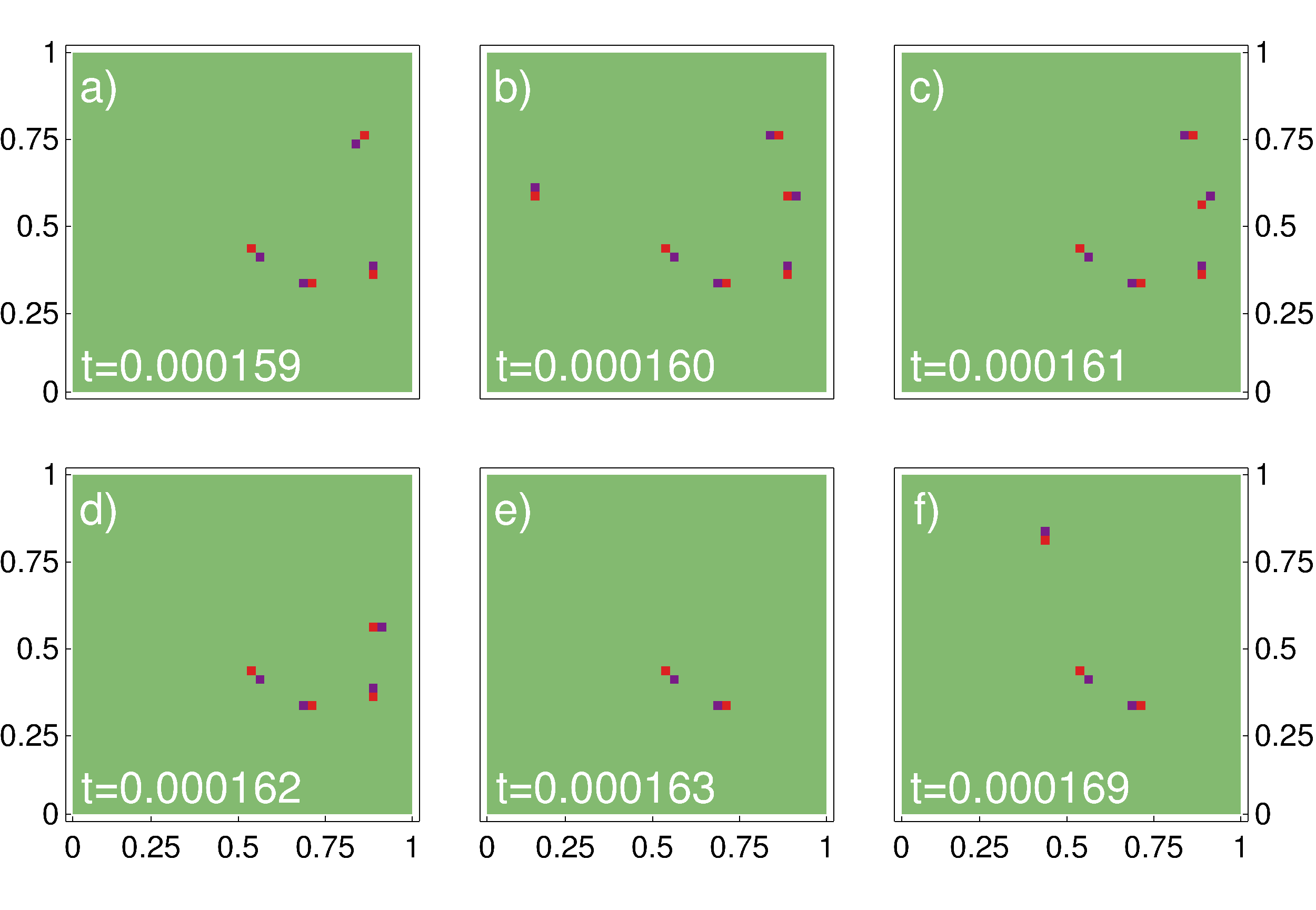}
\caption{Time-snapshots showing positions of virtual vortex-antivortex pairs in "+" component for a particular realization, for $\langle N_+\rangle=1500$, $\eta=3$, and at the temperature $T/T_F=0.1$. Time is scaled in the unit of $m L^2/\hbar$. Appearance and disappearance of vortex pairs on a time scale of $10^{-6}$ $m L^2/\hbar$ is clearly visible. }
\label{timesnapshots}
\end{figure} 

\begin{figure}[thb]
\center
\includegraphics[width=8.0cm]{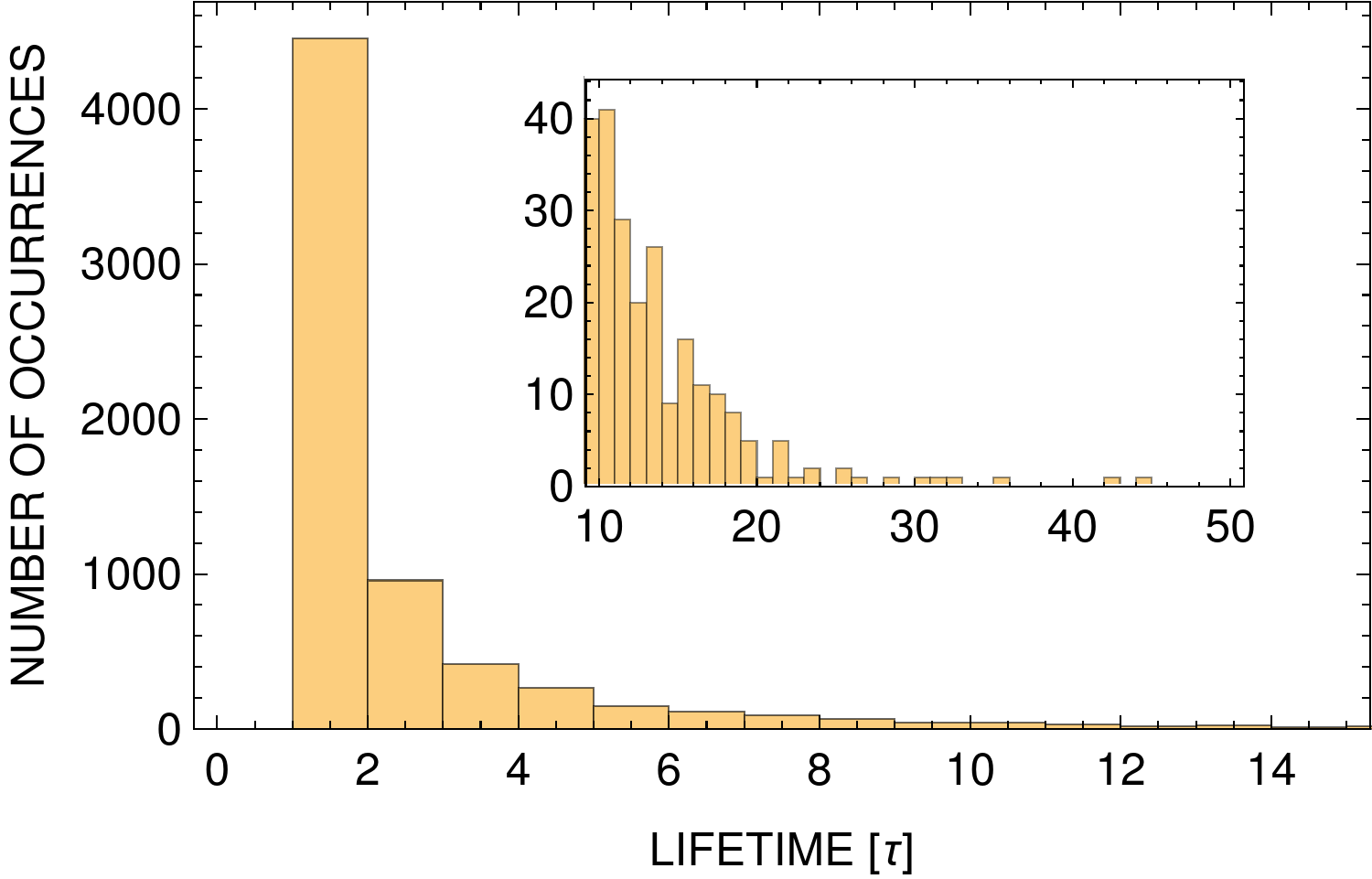}
\caption{Histogram of vortex-antivortex pairs lifetime. Each bin shows the number of occurrences of vortex pairs with a given lifetime, in units of $\tau=2 \times 10^{-6}$ $m L^2/\hbar$,  within a period of duration of $0.05$ $m L^2/\hbar$. Main plot shows the first $17$ bins (that corresponds to lifetimes $< 3.4\times 10^{-5}$ $m L^2/\hbar$, and the inset presents occupations of the rest of the bins ($< 10^{-4}$ $m L^2/\hbar$). This histogram was plotted based on the single realization studied in Fig. \ref{timesnapshots} but represents a typical outcome for a single real-time evolution of the system.}
\label{hist1}
\end{figure} 

\begin{figure}[thb]
\center
\includegraphics[width=8.0cm]{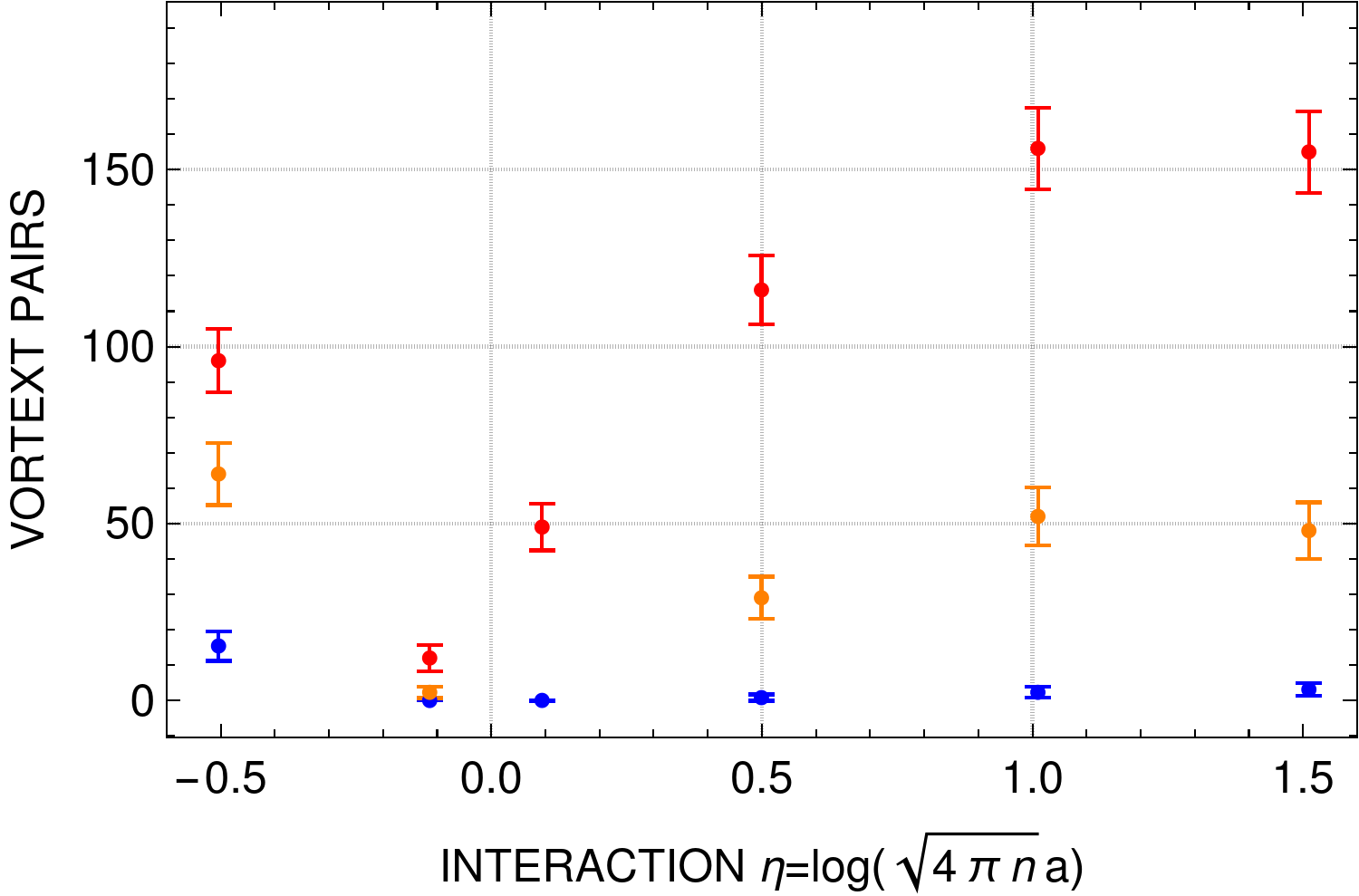}
\caption{Number of vortex pairs as a function of interaction parameter for different temperatures: $T/T_F=0.1$ (blue color), $T/T_F=0.15$ (orange color), and  $T/T_F=0.2$ (red color). Error bars represent the variance calculated based on three different sets of time evolution after imprinting the phase disturbed with $\phi_0/\pi=0.27,0.8,1.3$ for $T/T_F=0.1$, $\phi_0/\pi=0.8,1.3$ for $T/T_F=0.15$, and $\phi_0/\pi=1.3$ for $T/T_F=0.2$.}
\label{vpnumber}
\end{figure}

Strong damping of sound waves for positive and negative values of the interaction parameter is strictly related to the dimensionality of the system we study. In two-dimensional case the spectrum of elementary excitations becomes rich, qualitatively new modes appear with respect to what occurs in three-dimensional systems. Since the vortex energy depends logarithmically on the area occupied by the system, its creation is possible only at high enough temperatures, higher than the critical temperature for the Berezinskii-Kosterlitz-Thouless phase transition. However, for lower temperatures it is still energetically allowed to excite pairs of opposite charge vortices.

Our hypothesis is that the strong damping of sound waves traveling in a fermionic mixture is caused by the appearance of vortex-antivortex pairs in each fermionic component. Obviously, the vortices in the normal Fermi gas are not quantized (see Ref. \cite{Karpiuk03}) as they are in the superfluid phase. It is then expected that all fermions in a component, as described generally by individual orbitals, locally share the phase. Then locally each Eq. (\ref{equmotion2D}) becomes one to one correspondence to the set of the Hartree-Fock equations for a spin-polarized fermions (see Ref. \cite{Karpiuk20} for a derivation). The phase locally shared by individual fermions is just the local phase of the pseudo field while the density of the pseudo field is the sum of the densities of fermionic orbitals.

In Fig. \ref{timesnapshots} we show a sequence of time snapshots presenting positions of vortex-antivortex pairs in the "+" component while evolving a particular grand canonical microstate for $\langle N_+\rangle=1500$, $\eta=3$, and at the temperature $T/T_F=0.1$. From Fig. \ref{timesnapshots}, which demonstrates a typical dynamic behavior of fermionic components at nonzero temperatures, it is clear that vortex pairs appear on a time scale of $10^{-6}$ $mL^2/\hbar$ (which is a fraction of a microsecond for a typical size of the trapping box equal to $L\approx 40\,\mu$m \cite{Bohlen20}). Careful analysis of lifetimes of vortex-antivortex pairs leads to the histogram, Fig. \ref{hist1}. Since for low temperatures the vortex-antivortex correlations are straightforward and determined just by the distance between the vortex and antivortex, the evolution of pairs can be easily monitored. By looking at all possible separations from vortex to antivortex at each simulation time step, the vortex-antivortex pairs are those which correspond to the shortest distances (there is no ambiguity in this assignment because for low temperatures the number of vortex-antivortex pairs is small, see Fig. \ref{timesnapshots}). Since the energy of vortex pairs is of the order of $k_B T$, the product of their energy and a lifetime is of the order of $10^{-3} \hbar$, thus breaking the Heisenberg uncertainty principle. The vortex-antivortex pairs must be then the virtual pairs.

Hence, the attenuation of sound modes observed in the simulations is caused by the scattering on virtual vortex-antivortex pairs. The number of virtual vortex pairs is large for positive and negative $\eta$ (see Fig. \ref{vpnumber}) and the diffusivity is of the order of $\sim 10\, \hbar /m$, see Fig. \ref{figD} for temperatures $T/T_F>0.1$. Going to the strongly interacting regime the number of vortex pairs is significantly decreased and the attenuation coefficient gets smaller, for lower temperatures it takes values about  $1\, \hbar /m$, the universal quantum limit.

\begin{figure}[htb] 
\center
\includegraphics[width=7.0cm]{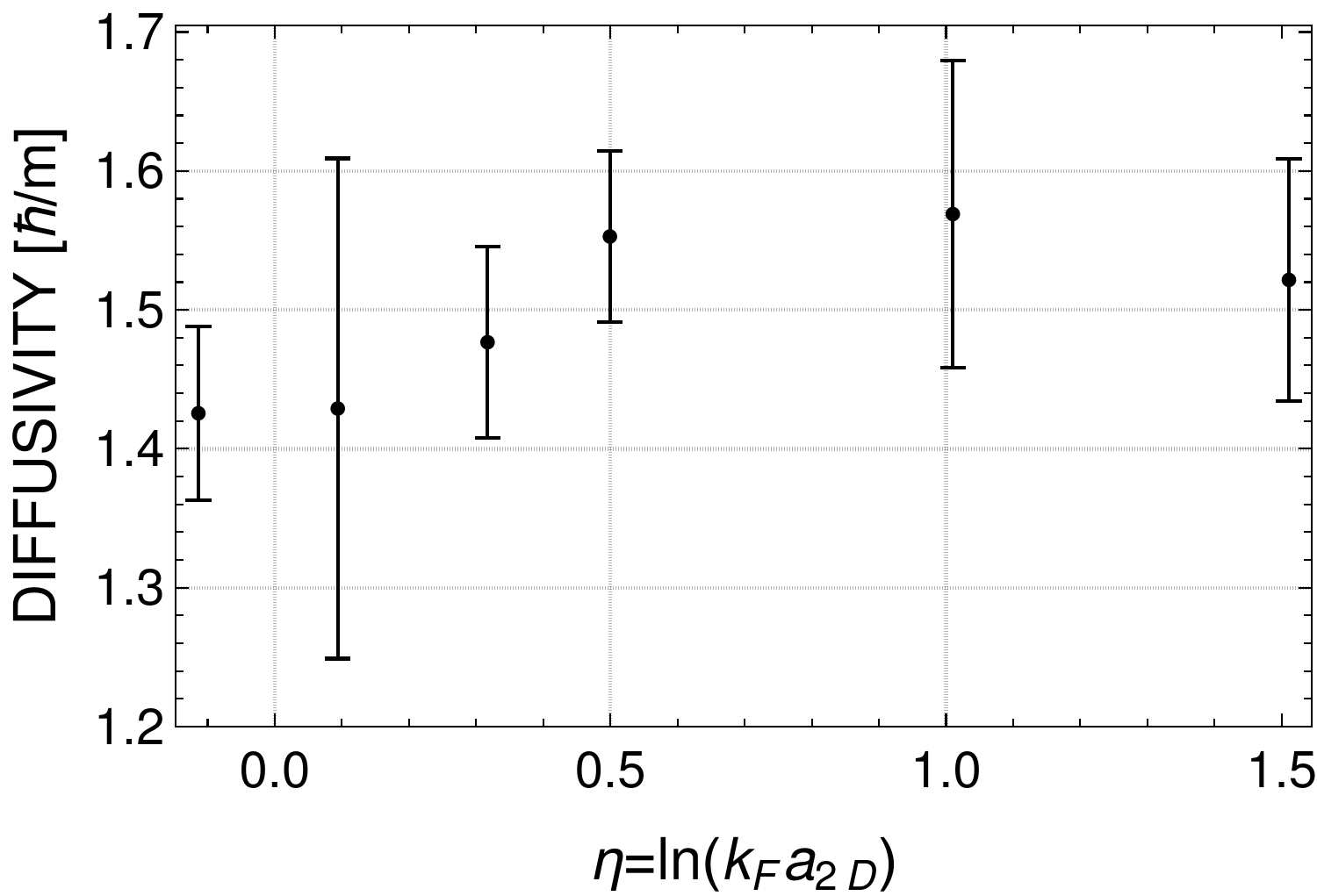} 
\caption{ Diffusivity as a function of interaction parameter for the temperature $T/T_F=0.2$ and for thermal fields containing no vortices. Error bars represent the variance calculated based on three different sets of time evolution after imprinting the phase $\phi_0 /\pi = 1.3$. The diffusivity, which stays close to the quantum limit, should be contrasted with huge values shown in Fig. \ref{figD} (red points).}  
\label{nodis}
\end{figure}

Finally, to uniquely indicate the origin of sound dissipation we generated an ensemble of microstates by enforcing constant phase for each pseudo-wave function $\psi_{\pm}$. The resulting thermal fields $\psi_{\pm}$ do not feature vortices but are still fluctuating density fields. After exciting a sound wave in a fermionic mixture we find that the sound propagates through a medium almost without damping, thus leading to diffusivity $D \approx 1\, \hbar /m$ close to the quantum limit both for positive and negative values of the interaction parameter, see Fig. \ref{nodis}. Hence, the presence of virtual vortex-antivortex pairs must be responsible for strong dissipation of sound waves. Similar conclusion can be found in Ref. \cite{Wu20}, where the authors discuss the propagation of sound waves in a two-dimensional ultracold bosonic gases by using a so called dynamic Kosterlitz-Thouless theory and deduce that the vortex dynamics is crucial to recover experimental data \cite{Ville18} on damping.

\section{Summary}
\label{conclusion}
In summary, we have studied the propagation of sound waves in a weakly and strongly interacting two-dimensional Fermi gas, in the range of temperatures close to the transition to a superfluid phase. We find numerically the dependence of the sound velocity and sound diffusivity on both the temperature and interactions. The sound diffusivity monotonically increases while the temperature is getting higher, independently of the interactions. At constant temperature the damping of sound takes the lowest values in the strongly interacting regime, where at temperatures close to the superfluid transition the diffusivity approaches the universal quantum limit. We identify that scattering on virtual vortex-antivortex pairs is responsible for strong dissipation of sound waves.

\section{Suplementary materials: LOCV approximation in two dimensions}
\label{locv2D}

The Hamiltonian of a uniform two-component two-dimensional (2D) degenerate Fermi gas is given by 

\begin{eqnarray}
H = - \frac{\hbar^2}{2 m} \sum_{i=1}^{N_+} \nabla_i^2 
      - \frac{\hbar^2}{2 m} \sum_{j=1}^{N_-} \nabla_j^2   
+ \sum_{i=1}^{N_+} \sum_{j=1}^{N_-} V_{FF}({\bf x}_i - {\bf y}_j)   \,,
\label{HFF}
\end{eqnarray}
where $V_{FF}$ is a zero-range pseudopotential that lead to the scattering length in two dimensions, $a_{2D}$. The many-body ground state of $H$ is assumed to be given in the form of the Jastrow-Slater variational wave function 
\begin{eqnarray}
|\Psi_{JS}\rangle = \prod_{i,j} f({\bf x}_i - {\bf y}_j)\,\,   |\Psi_S^+\rangle\,   |\Psi_S^-\rangle  \,,
\label{JSFF}
\end{eqnarray}
where $|\Psi_S^\pm\rangle$ is the Slater determinant wave function of a component consisting of $N_\pm$ fermions and $f({\bf r})$ is the Jastrow function describing the two-body correlations between interacting fermions. The pair correlation function, which is assumed to be spherically symmetric, is determined variationally by minimizing the average value of the energy $\langle \Psi_{JS}| H | \Psi_{JS} \rangle / \langle\Psi_{JS}|\Psi_{JS}\rangle$. Within the LOCV approximation the correlation function fulfills the Schr\"odinger equation
\begin{eqnarray}
-\frac{\hbar^2}{m} \left(\frac{d^2\!f}{dr^2} + \frac{1}{r} \frac{d\!f}{d r} \right) + V_{FF}(r)\, f(r) = \xi\, f(r) 
\label{SchFF2D}
\end{eqnarray}
in a space region defined by $r<d$. The healing length $d$, which is of the order of an average atomic separation, is determined self-consistently from the conditions $f(r > d) = 1$ and $f'(r = d) = 0$. For $r > d$, $f(r)$ tends to unity and the correlations disappear from (\ref{JSFF}). In the LOCV method the interaction energy $E_{int}/N$ is approximated by
\begin{eqnarray}
E_{int}/N = 2\pi n\,  \xi \int_0^d r\, |f(r)|^2\, dr 
\label{ENFF}
\end{eqnarray}
and $d$ is chosen such that
\begin{eqnarray}
2\pi n\,  \xi \int_0^d r\, |f(r)|^2\, dr = 1   \,,
\label{normFF}
\end{eqnarray}
where $n$ is the atomic density. Hence the interaction energy is $E_{int}/N = \xi$ (on average there are only correlated pairs). For the scattering state (the second lowest branch) $\xi = \hbar^2 k^2/ m > 0$ and in the noninteracting case the solution of Eq. (\ref{SchFF2D}) is
\begin{eqnarray}
f(r) \propto\,  a(k) J_0(k r) + b(k) Y_0(k r)  \,,
\label{LOCVsol}
\end{eqnarray}
where $a(k)$ and $b(k)$ are coefficients set by the boundary conditions and $J_0(k r)$ and $Y_0(k r)$ are Bessel functions of the first and second kinds, respectively. The two-dimensional contact interactions can be introduced by imposing the Bethe-Peierls boundary conditions at $r=0$
\begin{eqnarray}
\left( r \frac{d}{dr} - \frac{1}{\ln{(r/a_{2D})}} \right) f(r)  \underset{r \rightarrow 0}{\longrightarrow}  0   \,,
\label{BPbc}
\end{eqnarray} 
which gives \cite{Whitehead16}
\begin{eqnarray}
f(r) \propto\, \left\{ J_0(k r) - \frac{\pi}{2[\gamma + \ln{(k\, a_{2D}/2)}]} Y_0(k r) \right\}  \,,
\label{fsol}
\end{eqnarray}
where $\gamma \approx 0.577$ is the Euler’s constant. Since $f(r=d)=1$, the constant of proportionality equals
\begin{eqnarray}
\left\{ J_0(k d) - \frac{\pi}{2[\gamma + \ln{(k\, a_{2D}/2)}]} Y_0(k d) \right\}^{-1}  \,.
\label{propcon}
\end{eqnarray}
To determine the healing length ($d$) and the interaction energy per particle ($\sim k^2$), the constraint (\ref{normFF}) and the condition $f'(r = d) = 0$ have to be used. The second condition yields
\begin{eqnarray}
J_1(k d) = \frac{\pi}{2[\gamma + \ln{(k\, a_{2D}/2)}]}\, Y_1(k d)  \,.
\label{kd1}
\end{eqnarray}
Set of nonlinear equations (\ref{normFF}) and (\ref{kd1}) can be solved numerically giving quantities $(k,d)$ for each value of 2D interaction parameter $\eta = \ln{(k_F a_{2D})}$, where $k_F=\sqrt{4\pi n}$ is the Fermi momentum of a two-dimensional single-component Fermi gas of density $n$. Similar considerations can be repeated for the bound state (the lowest energy branch), just by making a replacement $k \rightarrow i \kappa$ (then $\xi = -\hbar^2 \kappa^2/ m < 0$).

\begin{figure}[thb] 
\center
\includegraphics[width=7.0cm]{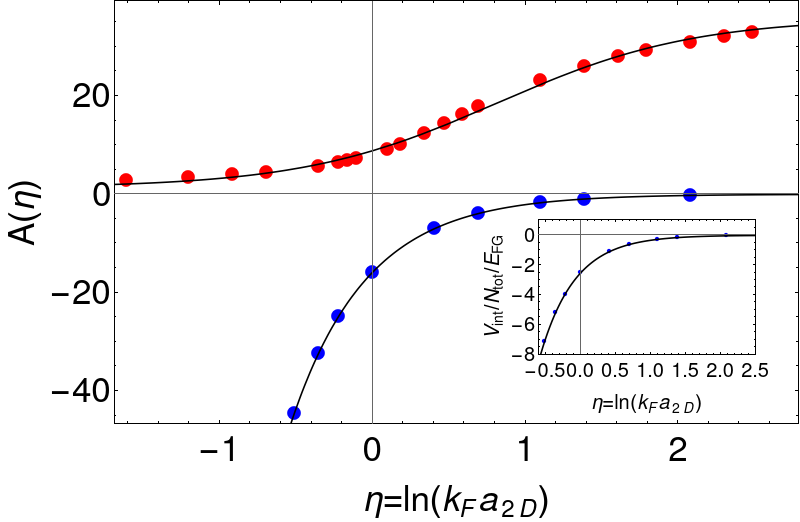}
\caption{$A(\eta)$, where $\eta = \ln{(k_F\, a_{2D})}$, calculated within 2D LOCV approximation for two lowest energy branches. The solid black lines are just a guide to the eye. Inset shows the interaction energy per particle in units of the energy per particle of the noninteracting gas, $V_{int}/N_{tot}/E_{FG}$, where $E_{FG}=\pi (\hbar^2/m) n $. The data compare well with the results of the fixed-node diffusion Monte Carlo calculations of Ref. \cite{Bertaina11} (see Fig. 1).}  
\label{Aeta}
\end{figure}

Now, the interaction energy density of each uniform component, calculated within the LOCV approximation, is $E_{int}^{\pm}/V=\xi n_\pm=4\pi (\hbar^2/m)\, n_\pm n_\mp (k/k_F^\mp)^2$. Hence, the total interaction energy density is 
\begin{equation}
	E_{int}/V=(E_{int}^{+}/V + E_{int}^{-}/V) /2 = (\hbar^2/2 m)\, n_+ n_- \left[A(\eta_+) + A(\eta_-)\right], 
\end{equation}
	where $A(\eta) = 4\pi (k/k_F)^2$. The function $A(\eta)$ is shown in Fig. \ref{Aeta} for two lowest energy branches. Applying the local density approximation the interaction energy functional of a two-component two-dimensional Fermi gas can be written as
\begin{eqnarray}
V_{int} = \int \frac{\hbar^2}{2 m}\, n_+({\bf r})\, n_-({\bf r}) \left[ A(\eta_+({\bf r})) + A(\eta_-({\bf r})) \right] d{\bf r}  \,.  
\label{LOCV2Dint}
\end{eqnarray}
Then the contribution to equations of motion, Eqs. (4) main text, due to interactions (the chemical potential $\mu_{int}$) becomes
\begin{eqnarray}
\frac{\delta V_{int}}{\delta n_\pm} = \frac{\hbar^2}{2 m}\, n_\mp \left[ A(\eta_+) + A(\eta_-) + \frac{1}{2} \frac{d A(\eta)}{d \eta} \Big{|}_{\eta_\pm}  \right]  
\label{chempot}
\end{eqnarray}
and the cutoff energy for a spin-balanced mixture consisted of $\langle N_{\pm} \rangle =1500$ atoms at the temperature $T/T_F=0.2$ as a function of 2D interaction parameter is plotted in Fig. \ref{cutoffenergy}. The inset depicts the chemical potential $\mu_{int}$.
\begin{figure}[thb] 
\center
\includegraphics[width=7.0cm]{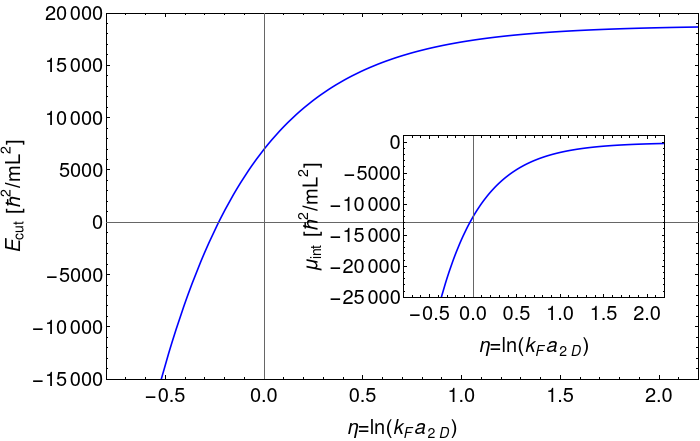}
\caption{Cutoff energy as a function of $\eta = \ln{(k_F\, a_{2D})}$ for the mixture consisted of $\langle N_{\pm} \rangle =1500$ atoms at the temperature $T/T_F=0.2$. Inset shows the chemical potential $\mu_{int}$. }  
\label{cutoffenergy}
\end{figure}

\section*{Acknowledgments}
Part of the results were obtained using computers at the Computer Center of University of Bialystok.

\section*{Availability of Data and Materials}
The datasets used and analysed during the current study available from the corresponding author on reasonable request.

\section*{Author contributions statement}
All authors made essential contributions to the work, discussed results, and contributed to the writing of the manuscript. The numerical simulations were performed by K.G.

\end{document}